\documentclass[letter,twocolumn]{jpsj2}

\title{Magnetic Susceptibility of Multiorbital Systems}

\author{Katsunori \textsc{Kubo} and
Takashi \textsc{Hotta}}
\inst{Advanced Science Research Center,
Japan Atomic Energy Research Institute,
Tokai, Ibaraki 319-1195}

\recdate{September 30, 2005; revised October 17, 2005;
  accepted October 25, 2005; published December 26, 2005}

\abst{
Effects of orbital degeneracy on magnetic susceptibility
in paramagnetic phases are investigated within a mean-field theory.
Under certain crystalline electric fields,
the magnetic moment consists of two independent moments,
\textit{e.g.}, spin and orbital moments.
In such a case, the magnetic susceptibility is given
by the sum of two different Curie-Weiss relations, 
leading to deviation from the Curie-Weiss law.
Such behavior may be observed in $d$- and $f$-electron systems
with $t_{2g}$ and $\Gamma_8$ ground states, respectively.
As a potential application of our theory, we attempt to
explain the difference in the temperature dependence of
magnetic susceptibilities of UO$_2$ and NpO$_2$.
}

\kword{
  Curie-Weiss law,
  orbital degeneracy,
  octupole moment,
  UO$_2$, NpO$_2$}

\begin{document}
\maketitle



The Curie-Weiss law is a well-known fundamental relation
that characterizes the magnetic response of condensed matter~\cite{Kittel}.
Indeed, in various kinds of materials, magnetic susceptibility $\chi$
is known to follow the Curie-Weiss law given by
\begin{equation}
  \chi=C/(T-\theta),
\end{equation}
where $T$ is the temperature, $C$ is the Curie constant,
and $\theta$ is the Weiss temperature.
The Curie-Weiss law is obtained by applying mean-field theory
to a model in which magnetic moments interact with one another
under an applied magnetic field.
In the mean-field approximation, $C \propto \mu_{\rm eff}^2$ and
$\theta=C\lambda$,
where $\mu_{\rm eff}$ is the effective magnetic moment
and $\lambda$ is the coupling constant of
the interaction between magnetic moments.
Thus, $\mu_{\rm eff}$ and $\lambda$ can be estimated by fitting
the experimental results for $\chi$ to the Curie-Weiss form.


On the other hand, there exist many materials in which the magnetic
susceptibility simply does not follow the Curie-Weiss law.
One reason for this, of course, is the effects of fluctuations,
which are not included in the mean-field approximation.
In order to improve the theory, one must take into account the effects of
fluctuations beyond the mean-field approximation in the temperature region
near a phase transition point.

However, even if the temperature is so high that fluctuation effects
can be ignored, the deviation from the Curie-Weiss law is still significant
in some materials with active \textit{orbital} degrees of freedom.
In such a case, the degeneracy of the ground-state multiplet is
generally lifted by the effects of a crystalline electric field (CEF).
When the interactions between magnetic moments are considered
within a mean-field theory,
we obtain the magnetic susceptibility as
\begin{equation}
  \chi^{-1}=\chi^{-1}_{\text{CEF}}-\lambda,
  \label{traditional_chi}
\end{equation}
where $\chi_{\text{CEF}}$ indicates the magnetic susceptibility
of a single ion in a CEF.
When the CEF level splitting between the ground and excited states
is much larger than thermal energy, it is enough to consider
only the CEF ground state to evaluate $\chi_{\text{CEF}}$.
In this case, since $\chi_{\text{CEF}}$ follows the Curie law, that is,
$\chi_{\text{CEF}}=C/T$, $\chi$ again follows the Curie-Weiss law.


From the above discussion, when the temperature is higher than
the phase transition point and the excited state is well separated from
the ground state, a Curie-Weiss law seems to hold,
but it is premature to conclude it.
An important point has been missed.
Thus far, we have implicitly assumed a one-component magnetic moment.
In some materials with an active orbital degree of freedom, however,
the magnetic moment consists of two components.
For instance, readers can envisage a $d$-electron system
in which orbital angular momentum is not quenched.
In such a system, the magnetic moment is given by
$\mib{M}=-\mu_{\text{B}} (2\mib{S}+\mib{L})$,
where $\mu_{\text{B}}$ is the Bohr magneton,
$\mib{S}$ is the spin angular momentum,
and $\mib{L}$ is the orbital angular momentum.

In this Letter, we study magnetic susceptibility in the paramagnetic
phase of a cubic system.
Note that in such a system with high symmetry, the orbital degree of freedom
becomes active in general, and the magnetic moment consists of
two independent moments in some cases.
It is intuitively understood that in such a situation,
the magnetic susceptibility is given by the sum of two different Curie-Weiss
relations, leading to non-Curie-Weiss behavior.
This is our main message,
but an explicit form of the magnetic susceptibility
of the system with two independent moments
is needed for the analysis of actual materials.
Thus, first we derive magnetic susceptibility
from a phenomenological model
in a level of standard mean-field approximation
by using simple algebra.
As examples, we consider
a $d$-electron system with a $t_{2g}$ ground state
and an $f$-electron system with a $\Gamma_8$ CEF ground state.
In particular, we compare our magnetic susceptibility with
the experimental results of NpO$_2$~\cite{Ross,Erdos,Aoki}
with a $\Gamma_8$ CEF ground state~\cite{Fournier,Amoretti}
and UO$_2$~\cite{Arrott,Homma}
with a $\Gamma_5$ CEF ground state~\cite{Kern,Amoretti2}.
The difference in the temperature dependence of magnetic susceptibilities
in these materials is naturally explained by considering the fact that
dipole \textit{and} octupole moments coexist in NpO$_2$, whereas
only the dipole moment exists in UO$_2$.


To consider both situations with and without spin-orbit interaction,
we generally express the magnetic moment $\mib{M}$ as
$\mib{M}=\mib{M}^{(1)}+\mib{M}^{(2)}$
in the CEF ground-state multiplet,
where the operators $\mib{M}^{(1)}$ and $\mib{M}^{(2)}$
are assumed to be orthogonal to each other,
namely tr$(M^{(1)}_{\alpha} M^{(2)}_{\alpha})=0$.
Here, $\alpha$ indicates a cartesian component
and tr$(\cdots)$ denotes the sum of expectation values
over the CEF ground states.
For simplicity, we further assume that a magnetic field $\mib{H}$
is applied along the $z$-direction.
The effective Hamiltonian, then, is given by
\begin{equation}
  \mathcal{H} = -H \sum_i M_{iz} 
  -\sum_{i\mu\alpha,j\nu\beta}
   I^{(\mu\nu)}_{i\alpha,j\beta}
   M^{(\mu)}_{i\alpha} M^{(\nu)}_{j\beta},
\end{equation}
where
$i$ and $j$ denote lattice sites,
and $I^{(\mu\nu)}_{i\alpha,j\beta}$ is the interaction between
$M^{(\mu)}_{i\alpha}$ and $M^{(\nu)}_{j\beta}$.
Note that the Hamiltonian should also include more complicated interaction
terms among the multipole moments, but in the mean-field approximation
for a paramagnetic phase up to $O(H^1)$, such terms vanish.
Thus, they are simply dropped at this stage.


Let us now apply the mean-field approximation to $\mathcal{H}$.
Since we are considering the magnetic susceptibility in the
paramagnetic phase, the site dependence can be suppressed,
that is, $M^{(\mu)}_{i \alpha}=M^{(\mu)}_{\alpha}$.
In addition, for $\alpha \neq \beta$,
we can show $\sum_{j} I^{(\mu \nu)}_{i \alpha, j \beta}=0$
by using a symmetry argument about a cubic lattice.
For example, when the vector connecting sites
$i$ and $j_{\pm}$ is given by $\mib{r}_{\pm}=(\pm x,y,z)$,
the relation $I^{(\mu, \nu)}_{i x, j_+ y}=-I^{(\mu, \nu)}_{i x, j_- y}$
holds in a cubic lattice,
indicating that $\sum_j I^{(\mu,\nu)}_{i x, j y}=0$.
Namely, the mean-field from $M^{(\mu)}_{\alpha}$ to
$M^{(\nu)}_{\beta}$ should be zero.
As a result, $\langle M^{(\mu)}_x \rangle=\langle M^{(\mu)}_y \rangle=0$
under a magnetic field $\mib{H}$ along the $z$-direction,
where $\langle \cdots \rangle$ denotes the expectation value.
Thus, the mean-field Hamiltonian for the paramagnetic state
up to $O(H^1)$ is expressed by
\begin{equation}
\begin{split}
  \mathcal{H}_{\text{MF}} = -H M_z
  -& [\lambda_1 M^{(1)}_z \langle M^{(1)}_z \rangle
     +\lambda_2 M^{(2)}_z \langle M^{(2)}_z \rangle],
   \label{MF_Hamiltonian} 
\end{split}
\end{equation}
where $\lambda_{\nu}=2\sum_{i,j}I^{(\nu \nu)}_{iz,jz}/N$
and $N$ is the number of sites.
Note that we have chosen the decomposition of the magnetic moment
so as to satisfy
$\lambda_{12}=2\sum_{i,j}I^{\text{(12)}}_{iz,jz}/N=0$.
Note also that in general, there are other moments
which couple with magnetic moments~\cite{Sakai},
but symmetry arguments tell us that
the mean-field from such moments cancels out
in the paramagnetic state,
as the cancellation of
the mean-field from $M^{(\mu)}_{\alpha}$ to $M^{(\nu)}_{\beta}$
for $\alpha \neq \beta$.
On the other hand, $M^{(1)}_z$ and $M^{(2)}_z$ have the same symmetry.
Thus, eq.~\eqref{MF_Hamiltonian} does not change
even if we include other interactions,
unless we consider spontaneous symmetry breaking.


The magnetic susceptibility for each moment, defined as
$\chi_{\nu} \equiv \text{d} \langle M^{(\nu)}_z \rangle / \text{d}H$
($\nu=1$ or 2), follows a Curie-Weiss law
independently as $\chi_{\nu}=C_{\nu}/(T-\theta_{\nu})$,
where $\theta_{\nu}=\lambda_{\nu} C_{\nu}$.
Note that the Curie constant $C_{\nu}$ is not determined here,
since we do not set the explicit form for $\mib{M}^{(\nu)}$.
Then, the magnetic susceptibility is given by $\chi=\chi_1+\chi_2$,
but it is useful to express it as
\begin{equation}
  \chi=C\frac{T+\theta-\theta_1-\theta_2}{(T-\theta_1)(T-\theta_2)},
  \label{modified_Curie-Weiss}
\end{equation}
where $C=C_1+C_2$ and $\theta=(C_1 \theta_1+C_2 \theta_2)/C$.
At high temperatures,
the magnetic susceptibility follows the Curie-Weiss law
$\chi=C/(T-\theta)$ asymptotically.
We note that $\theta$ plays the role of
the Weiss temperature for the high-temperature behavior.

\begin{figure}[t]
  \includegraphics[width=8cm]{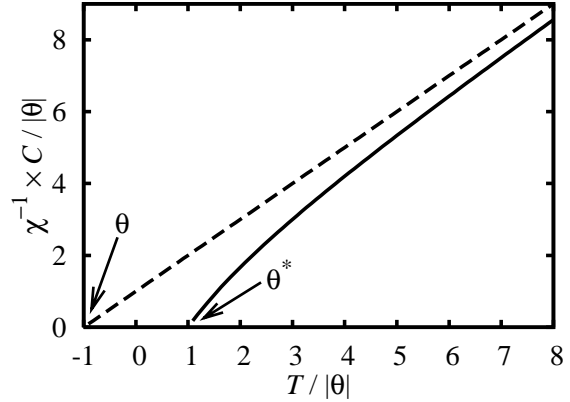}
  \caption{Temperature dependence of the inverse of the magnetic
    susceptibility $\chi$ (solid curve) obtained from
    eq.~\eqref{modified_Curie-Weiss}.
    The dashed line represents the Curie-Weiss behavior at
    high temperatures. The parameters are set as $C_1=C_2=C/2$,
    $\theta_1=|\theta|$, $\theta_2=-3|\theta|$, and $\theta < 0$.}
  \label{figure:chi_example}
\end{figure}

Figure~\ref{figure:chi_example} shows
a typical example of the temperature dependence of $\chi$.
First, we remark that $\chi$ diverges at a temperature $\theta^*$,
where $\theta^*$ is defined as the larger of $\theta_1$ and
$\theta_2$, and it is always larger than or equal to $\theta$.
We also note that $\theta^*$ can become positive,
even if $\theta$ is negative.
This fact indicates that if we naively assume a Curie-Weiss law,
we may obtain different Weiss temperatures.
In the worst case, one may mistake even the sign of the Weiss temperature,
which determines whether the interaction is ferromagnetic
or antiferromagnetic.
Thus, we should pay due attention to this effect
when analyzing the magnetic susceptibility of such systems.


Let us now discuss how our theory works in actual systems.
For this purpose, we choose two cases.
One is a $d$-electron system including only $t_{2g}$ ground
states, and the other is an $f$-electron system having a $\Gamma_8$
quartet ground state.


First, we consider a $t_{2g}$ system without the spin-orbit interaction.
There is no linear coupling between the spin and orbital angular momenta
due to the rotational symmetry in spin space.
Thus, we can simply decompose $M_z=M^{(1)}_z+M^{(2)}_z$ as
$M^{(1)}_z=-2\mu_{\text{B}} S_z$ and
$M^{(2)}_z=-\mu_{\text{B}} L_z$.
In the $t_{2g}$ subspace, the eigenvalues of the orbital angular
momentum $L_z$ and the spin angular momentum $S_z$ are given by
$\overline{L_z}=-1,0,1$ and $\overline{S_z}=\pm 1/2$, respectively,
where $\overline{A}$ denotes the eigenvalue of operator $A$.
Then, we obtain eq.~\eqref{modified_Curie-Weiss} with
$C_1=\mu^2_{\text{B}}/k_{\text{B}}$ and
$C_2=2\mu^2_{\text{B}}/(3 k_{\text{B}})$,
where $k_{\text{B}}$ is the Boltzmann constant.

Note that we have considered an ideal model
in order to see the role of the orbital angular momentum.
To include the effect of the spin-orbit interaction,
we should change the decomposition of $M_z$.
Moreover, in actual materials, the $t_{2g}$ level is easily
split into sublevels by Jahn-Teller distortions,
and the orbital angular momentum tends to be quenched.
Thus, the observation of the effects of the orbital angular momentum
on the magnetic susceptibility in $t_{2g}$ systems may be difficult.
Nevertheless, efforts to realize such a possibility are believed to
be challenging and interesting.


Next, we turn our attention to an $f$-electron system.
As is well known, due to the strong spin-orbit interaction,
the total angular momentum (dipole moment) $\mib{J}$,
rather than $\mib{S}$ or $\mib{L}$, is a good quantum number
in $f$-electron systems.
Thus, one is not allowed to decompose $\mib{M}$ into spin and
orbital angular moments as in the $t_{\rm 2g}$ electron system.
However, in $f$-electron systems, multipole moments higher than dipole
can influence the magnetic susceptibility under certain CEF potentials.
In fact, in a system on a cubic lattice with a $\Gamma_8$ CEF
ground state, the octupole moment $\mib{T}$ with the same symmetry
$\Gamma_{4u}$ as the dipole moment $\mib{J}$ becomes a degree of freedom
which is independent of the dipole moment~\cite{Shiina}.
Then, we can apply our theory to such a dipole-octupole system.
It is our message that the coexistence of dipole and octupole
moments can naturally lead to non-Curie-Weiss behavior
in some $f$-electron systems.

\begin{figure}[t]
\begin{center}
  \includegraphics[width=6cm]{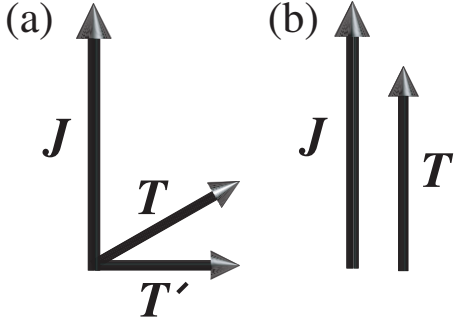}
  \caption{
    Schematic views for orthogonality between the dipole moment $\mib{J}$
    and $\Gamma_{4u}$ octupole moment $\mib{T}$.
    (a) In the $\Gamma_8$ subspace, $\mib{J}$ and $\mib{T}$
    are linearly independent, but not orthogonal to each other.
    Then, we define $\mib{T}^{\prime}$ by subtracting the component
    parallel to $\mib{J}$ from $\mib{T}$.
    (b) In the $\Gamma_{5}$ subspace, $\mib{T}$ is parallel to $\mib{J}$.
  }
  \label{figure:dipole_octupole_moments}
\end{center}
\end{figure}

We describe an effective model for the dipole-octupole system.
First, in the case of a strong spin-orbit interaction,
$\mib{M}$ is proportional to $\mib{J}$ through the well-known relation
$\mib{M}=-g_J \mu_{\text{B}} \mib{J}$,
where $g_J$ is Land\'e's $g$-factor.
Second, in the $\Gamma_8$ subspace,
we define an octupole moment $\mib{T}^{\prime}$
as a linear combination of the original $\Gamma_{4u}$ octupole moment
$\mib{T}$ and $\mib{J}$, in order that $\mib{T}^{\prime}$ is
orthogonal to $\mib{J}$
[see Fig.~\ref{figure:dipole_octupole_moments}(a)],
namely, tr$(T^{\prime}_{\alpha} J_{\alpha})=0$.
We normalize $\mib{T}^{\prime}$ so as to satisfy
$\text{tr}(J_{\alpha}J_{\alpha})
=\text{tr}(T^{\prime}_{\alpha}T^{\prime}_{\alpha})$.
In terms of $\mib{J}$ and $\mib{T}^{\prime}$,
the mean-field Hamiltonian is given by
\begin{equation}
  \begin{split}
    \mathcal{H}_{\text{MF}} = &-H M_{z}
    -(g_J \mu_{\text{B}})^2[\lambda_{\text{d}} J_z \langle J_z \rangle
    +\lambda_{\text{o}} T^{\prime}_z \langle T^{\prime}_z \rangle \\
    &+\lambda_{\text{do}}
    (J_z \langle T^{\prime}_z \rangle
    + T^{\prime}_z \langle J_z \rangle)],
  \end{split}
  \label{eq:G8model}
\end{equation}
where $\lambda_{\text{d}}$, $\lambda_{\text{o}}$,
and $\lambda_{\text{do}}$ are coupling constants of
dipole-dipole, octupole-octupole, and dipole-octupole interactions,
respectively.

Now, we define $M^{(1)}_z$ and $M^{(2)}_z$ so as to diagonalize
eq.~\eqref{eq:G8model} into the form of eq.~\eqref{MF_Hamiltonian}.
Then, we can exploit the present result for magnetic susceptibility,
indicating that $M^{(1)}_z$ and $M^{(2)}_z$ follow the Curie-Weiss law
independently.
We obtain eq.~\eqref{modified_Curie-Weiss} also
for $\Gamma_8$ quartet systems, with
$\theta$, $\theta_1$, and $\theta_2$
satisfying the relations
$\theta=C \lambda_{\rm d}$,
$\theta_1+\theta_2=C(\lambda_{\rm d}+\lambda_{\rm o})$,
and
$\theta_1\theta_2=
C^2(\lambda_{\rm d}\lambda_{\rm o}-\lambda_{\rm do}^2)$.
It has already been pointed out that the two $\Gamma_{4u}$ moments
follow the Curie-Weiss law with different Curie and Weiss constants
at least for certain interaction forms~\cite{Shiina},
but here we emphasize that this property is retained in general
when we choose the proper decomposition of $\mib{M}$.


As a typical application of eq.~\eqref{modified_Curie-Weiss}
to actual materials,
let us discuss the magnetic susceptibility of NpO$_2$~\cite{Ross,Erdos,Aoki},
in which the CEF ground state has been found to be
a $\Gamma_8$ quartet~\cite{Fournier,Amoretti}.
For comparison, we also consider the magnetic susceptibility
of UO$_2$~\cite{Arrott,Homma}
with a $\Gamma_5$ triplet CEF ground state~\cite{Kern,Amoretti2}.
Note that we can apply the present theory as long as
the CEF excitation energy is much higher than thermal energy.
In fact, the CEF excitation energies higher than a room temperature
were suggested for NpO$_2$~\cite{Fournier,Amoretti}
and UO$_2$~\cite{Kern,Amoretti2} from neutron scattering experiments.
The $\Gamma_8$ quartet has
a $\Gamma_{4u}$ octupole moment independent of the dipole moment
as mentioned before, and we expect non-Curie-Weiss behavior for NpO$_2$.
On the other hand, in the $\Gamma_5$ subspace,
the $\Gamma_{4u}$ octupole moment is proportional to
the dipole moment [see Fig.~\ref{figure:dipole_octupole_moments}(b)].
Thus, we expect simple Curie-Weiss behavior for UO$_2$.

\begin{figure}[t]
  \includegraphics[width=8cm]{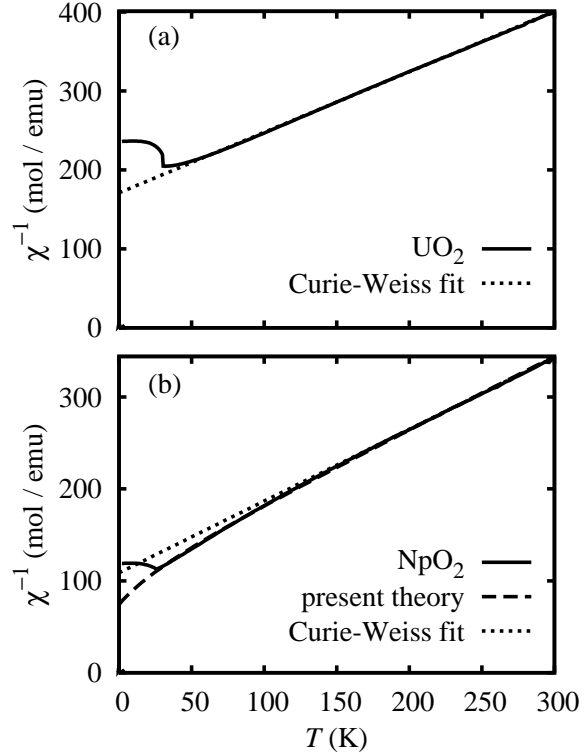}
  \caption{
    Temperature dependence of the inverse
    of the magnetic susceptibilities $\chi$ of UO$_2$ and of NpO$_2$.
    Solid curves indicate the experimental results for
    $H=10$~kOe and $\mib{H} \parallel [100]$
    (a) UO$_2$~\cite{Homma}.
    The dotted curve represents the Curie-Weiss law with
    $\mu_{\text{eff}}=3.23\mu_{\text{B}}$
    [$C=\mu^2_{\text{eff}}/(3k_{\text{B}})$]
    and $\theta=-223$~K.
    (b) NpO$_2$~\cite{Aoki}.
    The dashed curve represents the fit obtained with
    eq.~\eqref{modified_Curie-Weiss} with
    $\mu_{\text{eff}}=3.20 \mu_{\text{B}}$,
    $\theta=-146.1$~K, $\theta_{1}=-161.9$~K, and $\theta_2=-22.5$~K.
    The dotted curve represents the Curie-Weiss fit with
    $\mu_{\text{eff}}=3.20 \mu_{\text{B}}$ and $\theta=-139$~K.}
  \label{figure:UO2_NpO2}
\end{figure}

As shown in Fig.~\ref{figure:UO2_NpO2}(a),
the magnetic susceptibility of UO$_2$ follows the Curie-Weiss law
very well at $T \gtrsim 50$~K, at least up to 300~K.
The deviation from the Curie-Weiss law at $T \lesssim 50$~K
very likely originates from fluctuations relating to the
antiferromagnetic transition at 30.8~K.
On the other hand, in NpO$_2$ as shown in Fig.~\ref{figure:UO2_NpO2}(b),
the magnetic susceptibility deviates from the Curie-Weiss law below 150~K.
This temperature is much higher than the transition temperature 25~K,
and the deviation is unlikely to stem from critical phenomena.
In addition, the phase transition at 25~K in NpO$_2$ has been argued
to be due to ordering of $\Gamma_{5u}$
octupole moments~\cite{Paixao,Caciuffo,Lovesey,Kiss,Tokunaga,Sakai2,Kubo},
which do not influence the magnetic susceptibility in the paramagnetic phase.
As shown in Fig.~\ref{figure:UO2_NpO2}(b), the experimental data are
fitted well with eq~\eqref{modified_Curie-Weiss}.

Let us now try to explain the validity of the magnitude of parameters
obtained here.
For this purpose, it is useful to discuss coupling constants
among multipoles evaluated from the fitting parameters
$\theta$, $\theta_1$, and $\theta_2$.
After simple algebraic calculations, we obtain
$C \lambda_{\text{d}}=-145.1$~K,
$C \lambda_{\text{o}}=-38.3$~K, and
$C |\lambda_{\text{do}}|=44.2$~K.
It is found that the order of magnitude of 100~K of the dipole-dipole
interaction is consistent with the value of the spin-lattice
relaxation time $T_1$ deduced from $^{17}$O-NMR measurements
for NpO$_2$.~\cite{Tokunaga2}
On the other hand, concerning the magnitudes of octupole-octupole and
dipole-octupole interactions, it is difficult to deduce these values
from the experimental results.
However, for a system with active $\Gamma_{4u}$ octupole moments,
in general, $\lambda_{\text{o}}$ and $\lambda_{\text{do}}$
can take significant values.
In fact, from a microscopic $f$-electron model
on the basis of a $j$-$j$ coupling scheme,\cite{Kubo,Kubo2}
we have found that the magnitudes of $\lambda_{\text{o}}$ and
$\lambda_{\text{do}}$ are comparable to that of $\lambda_{\text{d}}$.
Thus, it seems reasonable to obtain the above values
for the coupling constants among multipoles for NpO$_2$.

It may be claimed that the experimental result for NpO$_2$
can be phenomenologically reproduced by $\chi=C/(T-\theta)+\chi_0$, where
$\chi_0$ denotes the so-called van Vleck term.
However, there seems no reasonable explanation why $\chi_0$ is significant
for NpO$_2$ whereas $\chi_0$ is almost zero for UO$_2$.
In addition, it is unclear how to divide the magnetic susceptibility into
``spin'' and ``orbital'' parts in $f$-electron systems with strong spin-orbit
coupling.
We claim that the difference in the temperature dependence of the magnetic
susceptibility between UO$_2$ and NpO$_2$ is naturally understood by
considering the effect of the $\Gamma_{4u}$ octupole moments in NpO$_2$.


In summary, we have discussed the effects of orbital degeneracy
on magnetic susceptibility in paramagnetic phases.
Under certain CEF potentials,
we have found that orbital or multipole moments coupled to
the magnetic moment can modify the Curie-Weiss law.
Other magnetic properties such as
nuclear magnetic resonance spectroscopy
will also be influenced by orbital or multipole moments.
For instance, on a simple cubic lattice with nearest-neighbor interactions,
the $z$-component of the octupole moment with $\Gamma_{5u}$ symmetry
couples to $J_z$ at a wave-vector $\mib{q}$
with $q_x \neq \pm q_y$~\cite{Sakai},
and the magnetic response at such wave-vectors will be influenced by
the $\Gamma_{5u}$ octupole moment even in the paramagnetic phase.


We are grateful to D. Aoki and Y. Homma for providing us with very recent
experimental results of the magnetic susceptibilities
in NpO$_2$ and UO$_2$, respectively, before publication.
We also thank R. H. Heffner, S. Kambe, N. Metoki, A. Nakamura,
H. Onishi, Y. Tokunaga, R. E. Walstedt, and H. Yasuoka
for fruitful discussions.
One of the authors (K. K.) is supported by the REIMEI Research Resources
of Japan Atomic Energy Research Institute.
The other author (T. H.) is supported by a Grant-in-Aid
for Scientific Research in Priority Area ``Skutterudites''
under contract No.~16037217 from the Ministry of
Education, Culture, Sports, Science, and Technology of Japan.
T. H. is also supported by a Grant-in-Aid for
Scientific Research (C)(2) under contract No.~50211496
from the Japan Society for the Promotion of Science.





\end{document}